\documentclass[superscriptaddress,aps,amsmath,amssymb,twocolumn]{revtex4-1}
\usepackage{times}
\usepackage[utf8x]{inputenc}
\usepackage[english]{babel}
\usepackage[T1]{fontenc}
\usepackage{lmodern}
\usepackage{bbold}
\usepackage[pdftex]{graphicx}
\usepackage{epstopdf}
\usepackage{bm}
\usepackage{color}
\usepackage{enumerate}

\newcommand{\bra}[1]{\langle#1|}
\newcommand{\ket}[1]{|#1\rangle}
\newcommand{\braket}[1]{\langle#1\rangle}
\newcommand{\mr}[1]{\mathrm{#1}}
\newcommand{\mc}[1]{\mathcal{#1}}
\newcommand{\ep}[1]{\langle#1\rangle}

\begin{document}
\title{Non-Markovian effect on quantum Otto engine: \\
-Role of system--reservoir interaction-}
\author{Yuji Shirai}
\affiliation{Department of Physics, The University of Tokyo, Komaba, Meguro, Tokyo 153-8505, Japan}
\author{Kazunari Hashimoto}
\email{hashimotok@yamanashi.ac.jp}
\affiliation{Graduate School of Interdisciplinary Research, University of Yamanashi, Kofu, Yamanashi 400-8511, Japan}
\author{Ryuta Tezuka}
\affiliation{Graduate School of Interdisciplinary Research, University of Yamanashi, Kofu, Yamanashi 400-8511, Japan}
\author{Chikako Uchiyama}
\email{hchikako@yamanashi.ac.jp}
\affiliation{Graduate School of Interdisciplinary Research, University of Yamanashi, Kofu, Yamanashi 400-8511, Japan}
\affiliation{National Institute of Informatics, Chiyoda, Tokyo,101-8430, Japan}
\author{Naomichi Hatano}
\email{hatano@iis.u-tokyo.ac.jp}
\affiliation{Institute of Industrial Science, The University of Tokyo, Kashiwa, Chiba 277-8574, Japan}%
\date{\today}

                         
\begin{abstract}
We study a limit cycle of a quantum Otto engine whose each cycle consists of two finite-time quantum isochoric (heating or cooling) processes and two quantum adiabatic work-extracting processes.
Considering a two-level system as a working substance that weakly interacts with two reservoirs comprising an infinite number of bosons, we investigate the non-Markovian effect (short-time behavior of the reduced dynamics in the quantum isochoric processes (QIPs)) on work extraction after infinite repetition of the cycles.
We focus on the parameter region where energy transferred to the reservoir can come back to the system in a short-time regime, which we call energy backflow to show partial quantum-mechanical reversibility.
As a situation completely different from  macroscopic thermodynamics, we find that the interaction energy is finite and negative by evaluating the average energy change of the reservoir during the QIPs by means of the full-counting statistics, corresponding to the two-point measurements.
The feature leads us to the following findings:
(1) the Carnot theorem is consistent with a definition of work including the interaction energy, although the commonly used definition of work excluding the interaction leads to a serious conflict with the thermodynamic law, 
and (2) the energy backflow can increase the work extraction.
Our findings show that we need to pay attention to the interaction energy in designing a quantum Otto engine operated in a finite time, which requires us to include the non-Markovian effect, even when the system-reservoir interaction is weak.

\end{abstract}

\pacs{Valid PACS appear here}
\maketitle\par

\section{Introduction}
The quantum heat engine (QHE) is becoming an important topic of interest from various perspectives: i) it is expected to retrieve and convert wasted heat in quantum devices into energy for work, which may thereby seed another industrial revolution; and ii) it may also offer a deeper understanding of thermodynamics from a quantum point of view.

The intensive studies stimulated by the first proposal of a QHE for the maser system~\cite{SSD} are typically classified as: (a) clarifying the thermodynamical laws and processes of a QHE by introducing the concept of open quantum dynamics~\cite{Alicki79,GK96,Kieu04,Quan05,Kieu06,Quan07,Quan09,Kosloff13,Seifert16,Li17}, (b) finding efficiency enhancements of heat engines using quantumness~\cite{Scully03,debate,Quan06,Gelbwaser15a,Niedenzu15,Turkpence16,Zhang07,Zhang08,Wang09,Dillenschneider,Thomas11,Altintas14,Hardal15,Doyeux16,Jaramillo16,Campisi16,Altintas16,Hardal18,Muller18,Latune19,Dambach19,Scully11,Huang12,Leggio16,Turkpence17,Li14,Gelbwaser15,Newman17,Newman20}, and (c) designing a finite-time operation of the QHE~\cite{Geva92,Feldmann96,Feldmann00,Feldmann03,Quan05,Rezek06,Wang12a,Wang12b,Wang13,Zheng16,Friedenberger17,Jeon17,Kosloff17,Funo}, including theoretical analysis on concrete experimental situations~\cite{Abah12,Rossnagel14,Rossnagel16,Fialko,Bergenfeldt,Zhang,Dong,Pekola16,Pekola18,Hardal17,Hofer16,Marchegiani16} and a report on a realization with NV centers~\cite{Klatzow19}.

In extending thermodynamics to the quantum regime, we face several challenges that have not been quite resolved in the literature.
First, open-system dynamics of a quantum working substance can be non-Markovian if the time scale of the working substance is comparable to or much shorter than that of the heat reservoirs, invalidating the application of the Gorini--Kossakowski--Sudarshan--Lindblad (GKSL) treatment~\cite{GKS,Lindblad} commonly used in conventional studies. This is because the treatment is legitimate only in the long-time limit, for which the correlation time of the coupling between working substance and reservoir is infinitesimally short~\cite{KTH,HSS,Breuer}. The limitation becomes critical when we design a heat engine with a finite operation time to obtain a finite power; note that, because of the infinite operational time, the Carnot engine with optimal thermal efficiency is practically useless, producing null power. Indeed, the non-Markovian effect has recently been intensively studied ~\cite{Breuer09,Breuer12,Angel10,Lu,Lorenzo,Luo,Chruscinski13, Angel14,Bylicka14,Addis,Guarnieri,Breuer16,Hashimoto19}.
One most characteristic feature is the backflow of information~\cite{Breuer09,Breuer12,Breuer16}, energy\cite{Guarnieri}, and spin~\cite{Hashimoto19} from the environment to the relevant system. The energy backflow, in particular, is found as a counterintuitive energy flow from a reservoir to a system which occurs in a very short time after a factorized initial condition.  The backflow occurs even if the effective temperature of a two-level system is equal to the reservoir temperature just before their interaction, reflecting the energy exchange between them to show partial reversibility, which we cannot obtain under the GKSL treatment~\cite{Guarnieri}. 

Second, the interaction energy between the working substance of the engine and the heat reservoir, which is ignored in classical thermodynamics because the dimensions are lower, plays a significant role in microscopic engines~\cite{Seifert16,Campisi11,Talkner20}.
Whether the contribution of the interaction Hamiltonian should be included into that for the working substance or reservoir remains controversial~\cite{Esposito15B}.
Because the interaction dynamically generates quantum correlations between the working substance and reservoir, we need to consider how to treat the interaction energy, for instance, as heat or work, especially when we take into account explicitly the attachment and detachment of the working substance to and from the reservoirs.

Third, the Carnot efficiency has been a distant unreachable limit for heat engines~\cite{Esposito10}. Whether we can exceed the limit by extending thermodynamics into the quantum regime has been a large motivation in the study of QHE.
All these issues demand a substantial updating of thermodynamics.

In this paper, we report our approach in addressing these aspects with a non-Markovian analysis regarding the finite-time operation of a quantum Otto engine (QOE), which comprises a two-level system and two bosonic reservoirs.
We use the time-convolutionless (TCL) non-Markovian master equation~\cite{Kubo63,Hanggi77,HSS,Shibata77,Chaturvedi79,Shibata80,US99,Breuer,Uchiyama14} to compute the dynamics of the two-level system during quantum isochoric processes (QIPs) with the work-extracting processes kept quantum adiabatic.

We notice a crucial role of the system--reservoir interaction in defining work for the non-Markovian QOE even in a weak-coupling regime.
While the energy cost of detaching the working substance from the reservoirs caused by the interaction is neglected in the conventional definition of work~\cite{Kieu04,Quan05,Kieu06,Quan07}, we find that the interaction takes a finite negative value in the non-Markovian dynamics, which indicates that we need to pay a cost for detachment against the attractive system--reservoir interaction.
By introducing a definition of work including the interaction, we show that one cannot extract positive work from the non-Markovian QOE operated beyond the Carnot efficiency, although an analysis based on the conventional definition excluding the interaction leads to a possibility of the positive work extraction.
This indicates that the thermodynamics law is consistent with the inclusion of the interaction to the work.
We also find that the energy backflow increases the amount of the extracted work to exhibit a maximum for a finite contact duration with reservoirs.

\section{Quantum Otto engine}
The QOE cycle consists of two quantum isochoric processes (QIPs) and two quantum adiabatic processes (QAPs)~\cite{Kieu04,Kieu06,Quan07}.
We describe it by a two-level system comprising a ``working substance'' attached to two reservoirs at different temperatures. Its total Hamiltonian is given by $\mathcal{H} = \mathcal{H}_{\mathrm{S}}(t) + \mathcal{H}_{\mathrm{B}} +\mathcal{H}_{\mathrm{I}}(t)$ with
\begin{align}
\mathcal{H}_{\mathrm{S}}(t) =\frac{\omega_{0}(t)}{2}\sigma_{z},\;\;\;\mathcal{H}_{\mathrm{I}}(t) = \sum_{\mu={\rm h,c}}\gamma_{\mu}(t) \mathcal{H}_{\mathrm{I}}^{\mu},
\end{align}
where we set \(\hbar=1\) and $\{\mu={\rm h}, {\rm c}\}$ labels the hot and cold reservoirs.
In $\mathcal{H}_{\mathrm{S}}(t)$, $\omega_{0}(t)$ denotes the Larmor frequency of the two-level system and $\sigma_{z}$ is the $z$-component of the Pauli matrix.
Since QIP is associated with the classical isochoric process in the sense that the working substance does not do any work during it, we keep $\omega_{0}(t)$ constant during each QIP~\cite{Quan07}.
Instead, we vary $\omega_{0}(t)$ quantum adiabatically in each QAP, meaning that we change it with the populations of the upper and the lower levels preserved.
In $\mathcal{H}_{\mathrm{I}}(t)$, $\gamma_{\mu}(t)$ is a time-dependent coefficient describing the contact switching of the $\mu$th reservoir.
Note that the present argument does not depend on the details of $\mathcal{H}_{\mathrm{I}}^{\mu}$ and $\mathcal{H}_{\mathrm{B}}$ until we perform numerical calculations for a specific case below. 

\begin{figure}[h]
\centering
\includegraphics[clip,width=1\columnwidth]{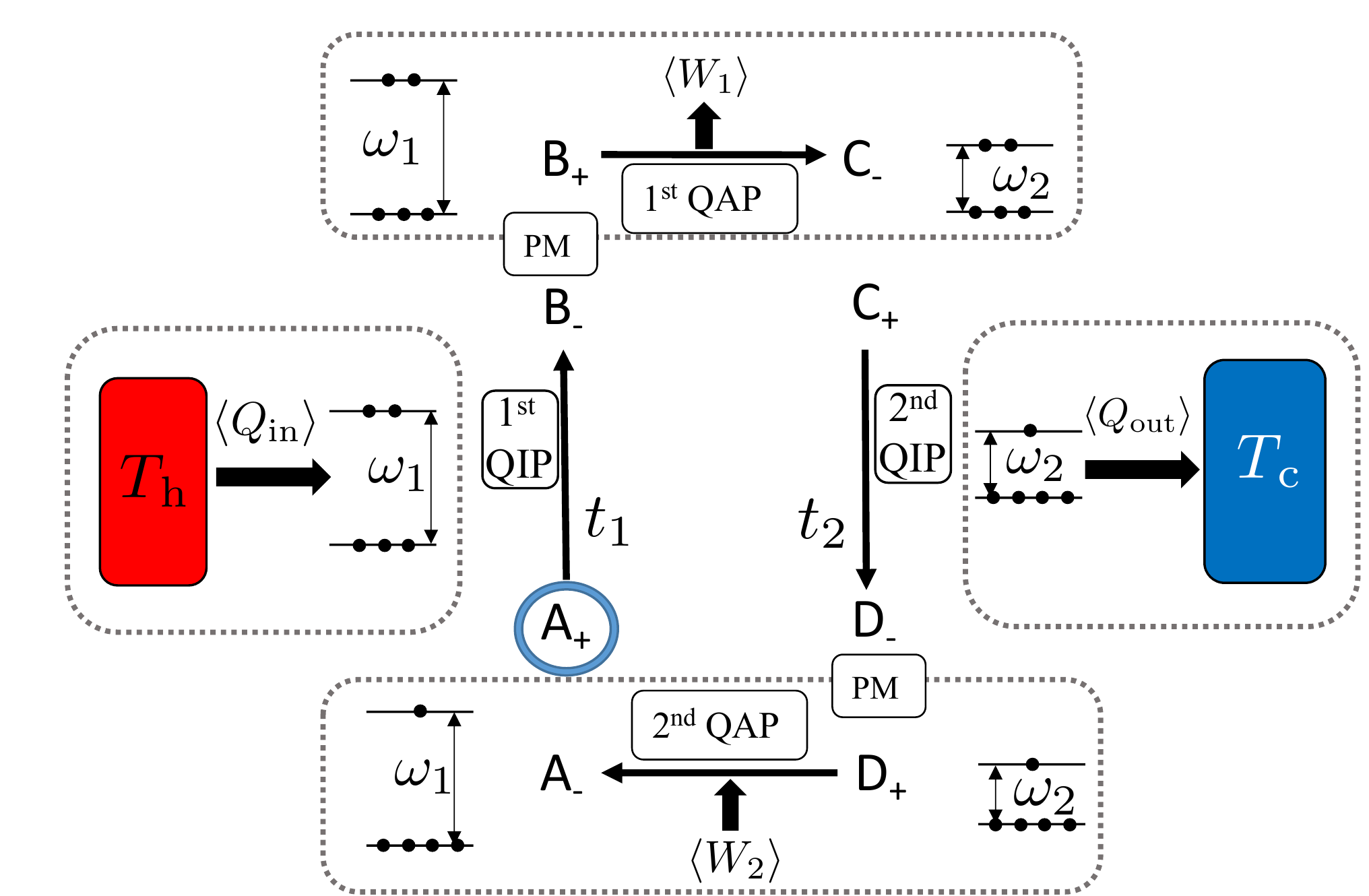}
\vspace{-10pt}
\caption{Our protocol of the QOE:
(1st QIP: $A_{+} \rightarrow B_{-}$) The system with the level difference $\omega_{\mathrm{h}}$ contacts the reservoir at $T_\mathrm{h}$ during $t_{1}$, exchanging energy. It is achieved by switching $\gamma_{\mathrm{h}}(t)$ to be set to unity and after time $t_1$ set back to zero, while the difference in level $\omega_{0}(t) = \omega_{\mathrm{h}}$ is maintained.
(1st QAP: $B_{+} \rightarrow C_{-}$) We decrease $\omega_0(t)$ quantum adiabatically from $\omega_{\mathrm{h}}$ to $\omega_{\mathrm{c}}$ maintaining the population of each level to transfer energy from the system to an external working storage, which may be realized by adiabatically expanding the box containing the two-level system.
(2nd QIP: $C_{+} \rightarrow D_{-}$) The system with $\omega_{\mathrm{c}}$ contacting the reservoir at $T_\mathrm{c}$ during $t_{2}$.
(2nd QAP: $D_{+} \rightarrow A_{-}$) We increase $\omega_{0}(t)$ quantum adiabatically from $\omega_{\mathrm{c}}$ back to $\omega_{\mathrm{h}}$, which may be realized by compressing the box by using energy stored in the external working storage.
The dots on the levels in the two-level system schematically represents the ratio of populations of the upper and the lower states.
}\label{fig:basic_cycle}
\end{figure}
We drive the system in the protocol shown in Fig.~\ref{fig:basic_cycle} in analogy to the classical Otto cycle.
In this protocol, we assume that the total system is prepared in a product state of the two-level system $\rho_{\rm{S}}(0)$ and the Gibbs state of the reservoir $\rho_{\rm{B}}^{\rm{h/c}}$, i.e. $\rho_{\rm{S}}(0)\otimes\rho_{\rm{B}}^{\rm{h/c}}$, before the system--reservoir contact is turned on, and the quantum correlations between the system and the reservoir are building up during each QIP through the interaction.
To ensure that the state of the total system reverts to a product state after each cycle, we introduce the projection measurement (PM)~\cite{ap61, ap62, ap63} after each QIP, severing quantum correlations and then use the resulting product state as the initial condition of the next cycle.
We assume that the reservoirs relax promptly to the Gibbs state after each PM in a much shorter time than one cycle, which determines the initial condition of the subsequent QIP.
For the present paper, we do not include any feedback based on information obtained from the measurements.\par

\section{Limit cycle}
We repeat the above cycle, assuming that it converges in the infinite-time limit, which we find here. Let $P^{\mu}_{n}$ denote the probability that the system is in the lower level $\ket{0}$ after the PM but before the $n$th contact with the $\mu$th reservoir. (We start counting the number of contacts before the attachment to the hot reservoir.)
They satisfy relations:
\begin{eqnarray}
\begin{split}\label{eq:probability}
&P^\mathrm{c}_{n}= P^\mathrm{h}_{n}\rho_{0,00}^{\mr{h}}(t_{1}) + (1-P^\mathrm{h}_{n})\rho_{1,00}^{\mr{h}}(t_{1}),\\
&P^\mathrm{h}_{n+1}= P^\mathrm{c}_{n}\rho_{0,00}^{\mr{c}}(t_{2}) + (1-P^\mathrm{c}_{n})\rho_{1,00}^{\mr{c}}(t_{2}),
\end{split}
\end{eqnarray} 
where $\rho_{m,\nu\nu}^{\mu}(t_{i})$ ($m,\nu=0,1$) denotes the $(\nu,\nu)$ element of $\rho^{\mu}_{m}(t_{i})$ representing the system density operator after contact for time $t_i$ with the $\mu$th reservoir under the factorized initial condition between the system $\ket{m}\bra{m}$ and the reservoir.
We assume that the $\mu$th reservoir is in the Gibbs state $\rho_{\mr{B}}^{\mu}$ with inverse temperature $\beta_{\mu}=1/T_{\mu}$; we set $k_{B}=1$ throughout this paper. In the limit $n\to\infty$, the probabilities converge to
\begin{align} 
P^{\mu} \equiv \lim_{n \rightarrow \infty}P^{\mu}_{n} 
= \frac{p^{\mu}}{1-p_0},\label{eq:Phc}
\end{align}
with
\begin{align}
&p_0\equiv \left[\rho_{0,00}^{\mr{c}}(t_{2}) - \rho_{1,00}^{\mr{c}}(t_{2})\right]\left[\rho_{0,00}^{\mr{h}}(t_{1})- \rho_{1,00}^{\mr{h}}(t_{1})\right],\\
&p^\mathrm{h/c}\equiv\rho_{0,00}^{\mr{c/h}}(t_{2/1})\rho_{1,00}^{\mr{h/c}}(t_{1/2})+\rho_{1,00}^{\mr{c/h}}(t_{2/1})\rho_{1,11}^{\mr{h/c}}(t_{1/2}).
\end{align}
We note that though $\rho^{\mu}_{m}(t_{i})$ depends on the dynamics during the QIP, the probabilistic relations Eqs.~\eqref{eq:probability} are valid whether the dynamics is Markovan or non-Markovian, and are independent of a specific form of the interaction Hamiltonian $\mathcal{H}_{\mathrm{I}}^{\mu}$.
Accordingly, the limit Eq.~\eqref{eq:Phc} is also quite general.
We hereafter analyze the QOE in this limit.

\section{Definitions of work}
In conventional studies of Markovian quantum heat engines in the weak-coupling regime, the widely used definitions of work and heat are based on the following separation of the change in the internal energy of the working substance:
$\left<dU(t)\right> = \left<dW(t)\right>+\left<dQ(t)\right>$, where the work done by an external force $\left<dW(t)\right>$ and the heat supplied from an external reservoir $\left<dQ(t)\right>$ are respectively defined as
	\begin{eqnarray}
	&&\braket{dW(t)}\equiv\mathrm{Tr}_{\mr{S}}\left[\rho(t)d\mc{H}_{\mr{S}}(t)\right],\label{eq:work0}\\
	&&\braket{dQ(t)}\equiv\mathrm{Tr}_{\mr{S}}\left[d\rho(t)\mc{H}_{\mr{S}}(t)\right],
	\end{eqnarray}
with $\mathrm{Tr}_{\mr{S}}$ denoting the partial-trace operation on the system~\cite{Kieu04,Quan05,Kieu06,Quan07}.
The definitions associate work and heat to the changes of the system Hamiltonian $d\mc{H}_{\mr{S}}(t)$ and of the state of the system $d\rho(t)$, respectively.
Specifically for the QOE, the system Hamiltonian changes only during QAPs. 
Since the 1st and 2nd QAPs correspond to expansion and compression processes, respectively, we denote the work done by the working substance during the 1st QAP as $W_{\rm{ad1}}$ and the work done by the external force to the working substance during the 2nd QAP as $W_{\rm{ad2}}$.
Using these quantities, the net amount of the work extracted from the engine during a single cycle is defined by
	\begin{equation}\label{eq:work1}
	W_{\rm{I}}\equiv W_{\rm{ad1}}-W_{\rm{ad2}}.
	\end{equation}
Because the population of the system is constant during each QAP, the energy changes of the system can be calculated as the difference between the system energies at the beginning and the end of each QAP.
Thus, the work done by the system during the 1st QAP is evaluated as
	\begin{equation}
	W_{\rm{ad1}}=(\omega_{\mathrm{h}}-\omega_{\mathrm{c}})\left[P^\mathrm{h}\rho_{0,11}^{\mr{h}}(t_{1})+(1-P^\mathrm{h})\rho_{1,11}^{\mr{h}}(t_{1})\right],\label{eq:work_1}
	\end{equation}
while the the work done by the external force to the system during the 2nd QAP is evaluated as
	\begin{equation}
	W_{\rm{ad2}}=(\omega_{\mathrm{h}}-\omega_{\mathrm{c}})\left[P^\mathrm{c}\rho_{0,11}^{\mr{c}}(t_{2})+(1-P^\mathrm{c})\rho_{1,11}^{\mr{c}}(t_{2})\right],\label{eq:work_2}
	\end{equation}
where $P^\mathrm{h/c}\rho_{0,11}^{\mr{h/c}}(t_{1/2})+(1-P^\mathrm{h/c})\rho_{1,11}^{\mr{h/c}}(t_{1/2})$ is the population of the excited state after the projection measurement taken place at the very end of each of the 1st and 2nd QIPs.
By using these expressions, we can derive the expression of the Otto efficiency in the form
	\begin{equation}
	\eta_{\mr{O}} = 1- \frac{\omega_{\mathrm{c}}}{\omega_{\mathrm{h}}},
	\label{eq:ottoeff}
	\end{equation}
relying on neither the Markovian approximation nor a specific form of system--reservoir interaction $\mathcal{H}_{\mathrm{I}}$.

Though the above definition of work is a reasonable extension of the classical first law of thermodynamics to quantum one, one might have doubt on its validity because of a crucial role of the system--reservoir interaction in quantum engines~\cite{Seifert16,Campisi11,Talkner20,Esposito15B}.
It has no classical counterpart, because in classical thermodynamics, the thermodynamic limit is taken on both of the system and reservoir, and thus the interaction energy is negligible.
In contrast, since the system remains small in quantum engines, the temporal change of the interaction energy during QIPs may not be negligible, and it can require us to pay a certain energy cost for the detachment of the system from the reservoir at the end of each QIP.

Such an insight leads us to define work by including the energy cost for detachment against the system--reservoir interaction.
To this end, we evaluate the expectation value of the interaction energy at the very end of each QIP, denoting $E_{\rm{I}}^{\mathrm{h}}(t_1)$ and $E_{\rm{I}}^{\mathrm{c}}(t_2)$.
If the interaction energy takes a negative value, we need a certain energy to detach the system against the attractive interaction, which results in a loss of the net amount of extracted work.
We thus define the work by
	\begin{equation}\label{eq:work2}
	W_{\rm{I\hspace{-.1em}I}}\equiv W_{\rm{I}}+E_{\rm{I}}^{\mathrm{h}}(t_1)+E_{\rm{I}}^{\mathrm{c}}(t_2).
	\end{equation}
Since the interaction energy is zero at the beginning of each QIP because of the factorized initial state, we can evaluate $E_{\rm{I}}^{\mathrm{h/c}}(t_{1/2})$ from the energy changes of the system and the reservoir during each of the 1st and 2nd QIPs, that is, $\Delta E_{\mathrm{S}}^{\mathrm{h/c}}(t_{1/2})\equiv\langle\mathcal{H}_{\mathrm{S}}^{\mr{h/c}}(t_{1/2})\rangle-\langle\mathcal{H}_{\mathrm{S}}^{\mr{h/c}}(0)\rangle$ and $\Delta E_{\mathrm{B}}^{\mathrm{h/c}}(t_{1/2})\equiv\langle\mathcal{H}_{\mathrm{B}}^{\mr{h/c}}(t_{1/2})\rangle-\langle\mathcal{H}_{\mathrm{B}}^{\mr{h/c}}(0)\rangle$, by using the energy conservation relation
	\begin{equation}\label{eq:interaction}
	E_{\mathrm{I}}^{\mathrm{h/c}}(t_{1/2})=-\Delta E_{\mathrm{S}}^{\mathrm{h/c}}(t_{1/2})-\Delta E_{\mathrm{B}}^{\mathrm{h/c}}(t_{1/2}).
	\end{equation}
In the expression, we can evaluate the quantities in the right-hand side as follows:
the energy change of the system during each QIP can be directly evaluated from the difference of the mean values of the system energy as $\Delta E_{\mathrm{S}}^{\mathrm{h/c}}(t_{1/2})={\rm Tr}_{\rm{S}}[\mathcal{H}_{\mathrm{S}}^{\mr{h/c}}(t_{1/2})\rho^{\mr{h/c}}(t_{1/2})]-{\rm Tr}_{\rm{S}}[\mathcal{H}_{\mathrm{S}}^{\mr{h/c}}(0)\rho^{\mr{h/c}}(0)]$;
instead, the energy change of the reservoir can be evaluated by using  full-counting statistics based on two successive projective measurements of the reservoir energy $\mathcal{H}_{\mathrm{B}}^{\mathrm{h/c}}$ \cite{Esposito} performed at the very beginning and very end of each QIP, whose derivation and expressions are summarized in Appendix~\ref{appC}.

\section{Numerical evaluation}
We now examine the validity of the definition of work Eq.~\eqref{eq:work2} by numerically analyzing work extraction of the engine.

\subsection{Model}
As a working system, we suppose that the two-level system interacts with reservoirs each of which consists of an infinite number of bosons described by the Hamiltonian
\begin{align}
\mathcal{H}_{\mathrm{B}} = \sum_{\mu={\rm h,c}}\mathcal{H}_{\mathrm{B}}^{\mu}=\sum_{\mu}\sum_{k}\epsilon_{k,\mu}b_{k,\mu}^{\dagger}b_{k,\mu}, 
\end{align}
where $b_{k,\mu}^\dagger$ and $b_{k,\mu}$ denote the creation and annihilation operators of mode $k$ of the $\mu$th reservoir, which has energy $\epsilon_{k,\mu}$.
Defining the interaction Hamiltonian as
\begin{align}
\mathcal{H}_{\mathrm{I}}^{\mu}= \sigma_{x}\otimes \sum_{k} \left(g_{k,\mu}b_{k,\mu}^{\dagger} + g_{k,\mu}^{*}b_{k,\mu}\right),
\end{align}
with interaction strength $g_{k,\mu}$ between the two-level system and the bosons of the $k$th mode of the $\mu$th reservoir, we evaluated the reduced dynamics during the QIPs by adopting the TCL master equation to describe the non-Markovian dynamics of the two-level system in contact with a reservoir.
For the model, the TCL master equation to second order is exactly solvable when the system--reservoir coupling is described by an Ohmic spectral density $J(\omega)\equiv\sum_k|g_k|^2\delta(\omega-\epsilon_k) = \lambda\omega \exp(-\omega/\Omega)$, where $\lambda$ is a coupling constant and $\Omega$ is a cutoff frequency~\cite{Guarnieri,Uchiyama14}; an explicit expression of the solution is given in Appendix~\ref{appA}.

We note that the Born--Markov approximation is valid if the autocorrelation function of the bosonic reservoir in the TCL master equation decays much faster than the relaxation time of the two-level system through the system--reservoir interaction, which is achieved by taking the long-time (Markovian) limit $t\to\infty$ on the TCL master equation, Eq.~\eqref{eq:TCL}. It produces the Markovian master equation, which is exactly solvable with the solutions Eqs.~\eqref{eq:Msol1} and \eqref{eq:Msol2}.   Alternatively, the correlation time of the bosonic reservoir becomes shorter by setting simultaneously the coupling constant $\lambda$ sufficiently small and the cutoff frequency $\Omega$ large enough, corresponding to the Markovian approximation.
Since we focus on the non-Markovian effect on the heat engine in the present paper, we set $\Omega$ small to make the non-Markovian effect significant in the following numerical calulculations. (See Ref.~\cite{Guarnieri} for details of the dependence of the energy backflow on the cutoff frequency in the weak-coupling situation. Similar parameter setting was used to study a non-Markovianity measure in Ref.~\cite{clos12}.)

\subsection{Interaction Energy}
For the model specified above, we examine the time evolution of the interaction energy during the contact with the hot reservoir (1st QIP) after the QOE reach the limit cycle, by using Eq.~\eqref{eq:interaction}.
For this purpose, we evaluate the time evolution of the energy of the hot reservoir, $\Delta E_{\mathrm{B}}^{\mathrm{h}}(t)$, as well as of the system, $\Delta E_{\mathrm{S}}^{\mathrm{h}}(t)$, for $0 \le t \le t_{1}$, keeping the contact duration with the cold reservoir $t_{2}$ constant.

\begin{figure}[h]
\centering
\includegraphics[clip,width=1\columnwidth]{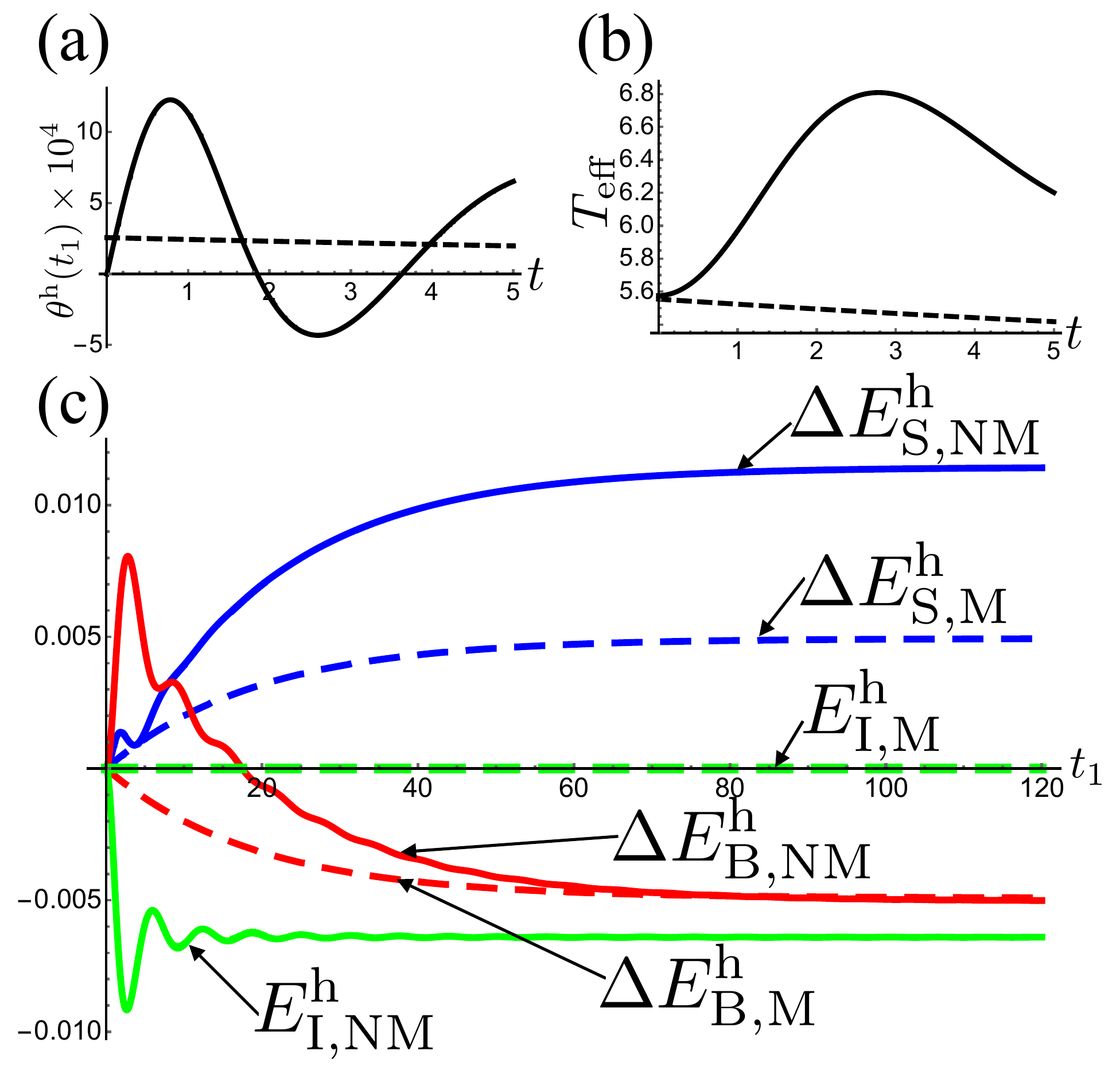}
\vspace{-10pt}
\caption{
Non-Markovian (NM; solid line) and Markovian (M; dashed line) dynamics of (a) the energy flow, (b) the effective temperature, and (c) the change in energy respective parts of the engine during contact with a hot reservoir; a flow to the reservoir is defined to be positive: We find that (a) the non-Markovian dynamics shows the energy backflow from the reservoir and (b) the effective temperature of the system becomes higher than the hot reservoir. In (c), we show the change in energy, 
$\Delta E_{\mr{S}}^{\mr{h}}$, $\Delta E_{\mr{B}}^{\mr{h}}$ and $E_{\mr{I}}^{\mr{h}}$. We find that $E_{\mr{I}}^{\mr{h}}$ is finite in the non-Markovian dynamics.
The parameter settings are: $T_{\mr{h}} = 5.0$, $T_{\mr{c}} = 1.0$, $t_{2}=60$, $\lambda = 0.01$, $\Omega = 0.4$, and $\omega_{\mathrm{c}}/\omega_{\mathrm{h}} = 0.18$ for unit values $\omega_{\mathrm{h}} = 1$ with $k_\mathrm{B}=\hbar=1$.
If an actual value of the level splitting of the two-level system in contact with the hotter reservoir is $\hbar\omega_{\rm{h}}=1{\rm{meV}}$, the actual values of the other parameters are roughly evaluated as follows: $\lambda=10{\rm{meV}}$, $\hbar\Omega=0.4{\rm{meV}}$, $T_{\rm{h}}\approx58{\rm{K}}$, $T_{\rm{c}}\approx11.8{\rm{K}}$ and $\hbar\omega_{\rm{c}}=0.18{\rm{meV}}$.
}\label{fig:energy_changes}
\end{figure}

To investigate the feature of $\Delta E_{\mathrm{B}}^{\mathrm{h}}(t)$, let us firstly show its time differential coefficient, $d\Delta E_{\mathrm{B}}^{\mathrm{h}}(t)/dt (\equiv \theta^{\mr{h}}(t))$, which we call energy flow, whose sign represents the direction of the net energy transfer between the system and the hot reservoir. 
We define the energy flow to be positive, $\theta^{\mr{h}}(t)>0$, when the energy is flowing into the reservoir, and to be negative, $\theta^{\mr{h}}(t)<0$, when the energy is flowing into the system. 
In Fig.~\ref{fig:energy_changes}(a), we provide a numerical estimate of $\theta^{\mr{h}}(t)$ in the short-time duration $0\leq t\leq t_{1}$ for the non-Markovian (solid line) and Markovian (dashed line) cases after the QOE reach the limit cycle with the temperatures of the reservoirs set to $ T_{\mr{h}} = 5.0$ and $T_{\mr{c}} = 1.0$, the contact duration to $t_{1}=5$ and $t_{2}=60$, the coupling strength of system-reservoir interaction to $\lambda = 0.01$, the cutoff frequency of the spectral density to $\Omega = 0.4$, and the ratio of the Larmor frequencies of the system to $\omega_{\mathrm{c}}/\omega_{\mathrm{h}} = 0.18$ for the unit values $\omega_{\mathrm{h}} = 1$ with $k_\mathrm{B}=\hbar=1$.
The dashed line in Fig.~\ref{fig:energy_changes}(a) indicates that the energy always flows from the system into the reservoir in the Markovian case.
In the non-Markovian case, by contrast, the energy temporarily flows in reverse, which we call the energy backflow \cite{Guarnieri}.

In Fig.~\ref{fig:energy_changes} (b), we show the time evolution of the system in accordance with the above energy transfer by evaluating the temporal change of the ratio of populations of the upper state $\rho_{11}$ and of the lower state $\rho_{00}$,  $\omega_{\mathrm{h}}[\ln(\rho_{00}/\rho_{11})]^{-1} (\equiv T_{\mathrm{eff}})$, which corresponds to the effective temperature of the system.
While the Markovian case (dashed line) shows a monotonic approach to the population given in the equilibrium with the hot reservoir of temperature $ T_{\mr{h}} = 5.0$,  we find for the non-Markovian case (solid line) the population in the excited state becomes larger as indicated by the increase of the effective temperature of the system, corresponding to the energy backflow.  

We next present in Fig.~\ref{fig:energy_changes}(c) how the short-time behaviors of the energy transfer and the system affect the energy change of the respective parts of the engine, with changing $t_{1}$ while keeping other parameters the same as above for each case.
Let us first show a crucial role of the interaction energy in the non-Markovian dynamics in comparison with the Markovian dynamics:
whereas the energy transfer to the system--reservoir interaction is constantly zero in the Markovian case (see $E_{\rm{I,M}}^{\rm{h}}$ indicated by the green dashed line), it takes a negative finite value in the non-Markovian case (see $E_{\rm{I,NM}}^{\rm{h}}$ indicated by the green solid line).
This means that, in addition to the backflow, the system also withdraw energy from the system--reservoir interaction in the non-Markovian case, which causes attractive interaction between system and reservoir.
We thus need to take into account the interaction energy as a part of work when we detach the reservoir from the system.

\subsection{Work extraction}
\begin{figure}
\centering
\includegraphics[clip,width=1\columnwidth]{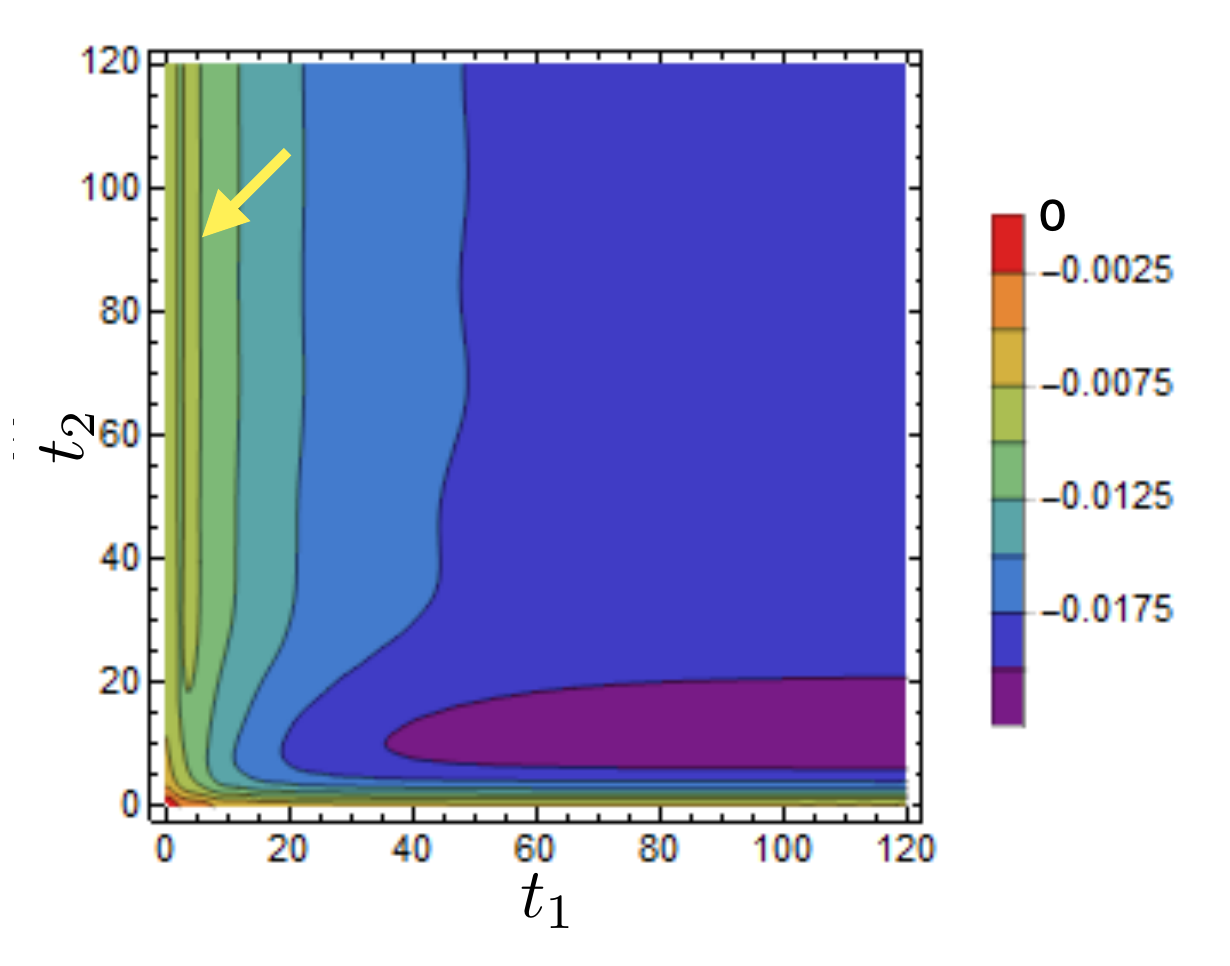}
\vspace{-10pt}
\caption{\label{fig:work_extraction_detach}
Extracted work $W_{\rm{II}}$, Eq.~\eqref{eq:work2}, of the QOE under the non-Markovian dynamics for contact durations $t_{1}$ and $t_{2}$ with the detachment energy taken into account.
The parameter settings are: $T_{\mr{h}} = 5.0$, $T_{\mr{c}} = 1.0$, $\lambda = 0.01$, $\Omega = 0.4$, and $\omega_{\mathrm{c}}/\omega_{\mathrm{h}} = 0.18$ for unit values $\omega_{\mathrm{h}} = 1$ with $k_\mathrm{B}=\hbar=1$ (same as in Fig.~\ref{fig:energy_changes}).
} 
\end{figure}
The negative value of the interaction in the non-Markovian dynamics (green solid line in Fig.~\ref{fig:energy_changes}) leads to the consideration that we need energy to detach the system from the reservoir against the attractive system--reservoir interaction; thus a part of applied work is consumed in the detachment, which is exactly the idea of our definition of work formulated by Eq.~\eqref{eq:work2}.
In Fig.~\ref{fig:work_extraction_detach}, we show the work extraction with changing $t_{1}$ and $t_{2}$ by numerically evaluating $W_{\rm II}$.
In the numerical calculations, we chose parameters such that the Otto efficiency $\eta_{\rm{O}}$ in Eq.~\eqref{eq:ottoeff} exceeds the Carnot efficiency $\eta_{\rm{C}}\equiv1-T_{\rm c}/T_{\rm h}$, i.e. $\eta_{\rm{O}}>\eta_{\rm{C}}$.
We now see that the total amount of work is negative for the entire region.
In contrast, the conventional definition of work formulated by Eq.~\eqref{eq:work1} provides a serious conflict with the thermodynamic law; $W_{\rm{I}}$ eventually becomes positive for the same parameter values with $\eta_{\rm{O}}>\eta_{\rm{C}}$ (see Appendix~\ref{app:W1} for details).
Hence, for the non-Markovian QOE, the thermodynamic law seems to be consistent with the inclusion of the interaction energy to the work.

Figure~\ref{fig:work_extraction_detach} also shows that the extracted work exhibits a maximum value for a finite set of $t_1$ and $t_2$ in the green region indicated by the yellow allow.
This is caused by the energy backflow;
as discussed in the previous section, the energy eventually flows from the reservoir to the system in the non-Markovian dynamics (the solid line in Fig.~\ref{fig:energy_changes}(a));
accordingly, the population of the excited state increases (the solid line in Fig.~\ref{fig:energy_changes}(b));
the increase of the population of the excited state contributes to the increase of $W_{\rm{ad1}}$, resulting in the maximum of the work extraction.
The maximum is another characteristics of the non-Markovian QOE.
Indeed, for the Markovian QOE, the amount of work extraction monotonically decreases with respect to $t_1$ and $t_2$ as shown in Fig.~\ref{fig2} in Appendix~\ref{appB}.
We confirm the conclusions above for several other values of $\omega_{\mathrm{h}}$ and $\omega_{\mathrm{c}}$ in Appendix~\ref{appD}.

\section{Conclusions and discussion}
We have examined the role of the system--reservoir interaction in the non-Markovian quantum Otto engine.
While the energy cost of detaching the system from the reservoir caused by the interaction is neglected in the conventional definition of the extracted work in quantum heat engine, we find that the energy of the system--reservoir interaction temporally changes to be negative during each quantum isochoric process in the non-Markovian quantum Otto engine.
Resulting attractive system--reservoir interaction requires us to pay a certain energetic cost to detach it at the end of each quantum isochoric process, thus reducing the net amount of extracted work.
By introducing a new definition of work including the interaction energy, we show that the net amount of extracted work remains negative if the parameters are chosen such that the Otto efficiency $\eta_{\rm{O}}$ exceeds the Carnot efficiency $\eta_{\rm{C}}$.
In contrast, we show in Appendix~\ref{app:W1} that the conventional definition of work excluding the interaction energy eventually becomes positive for the same parameter values with $\eta_{\rm{O}}>\eta_{\rm{C}}$.
This indicates that the thermodynamic law seems to be consistent with the inclusion of the interaction energy to the work in the non-Markovian quantum Otto engine.
We also find that the work exhibits a maximum for finite $t_1$ and $t_2$ due to the energy backflow.
The latter finding may be useful to design a highly efficient quantum heat engine possessing a finite power.

The above summarized features are characteristics of the non-Markovian engine.
Indeed, the numerical result presented in Fig.~\ref{fig:energy_changes}(c) shows that the interaction energy is constantly zero in the Markovian case; thus the definitions of work Eqs.~\eqref{eq:work1} and \eqref{eq:work2} coincide in the Markovian engine.
In Appendix~\ref{appB}, we summarized the work extraction in the Markovian engine.
Figure~\ref{fig2} in the Appendix shows that the work extraction from the Markovian engine operated under the condition $\eta_{\rm O}>\eta_{\rm C}$ is entirely negative, and the net amount of extracted work monotonically decreases as the contact durations $t_{1}$ and $t_{2}$ increases.

Regarding the treatment of the interaction energy in defining work and heat, the controversy has been resolved for strongly coupled systems~\cite{Campisi11,Esposito15B,Seifert16,Talkner20}.
In the present work, however, we found that the finite contribution of the interaction energy under the non-Markovian effect is relevant even for weak system--reservoir coupling.
The finding dictates a reconsideration of the foundation of the controversy.
In terms of the work-extracting procedure, several studies \cite{Watanabe,Paz14,ap72,Paz15,Talkner16,ap64,ap76} suggest that we may need to consider a coupling between the system and measurement apparatus.
Further work is necessary to make the present QOE experimentally feasible.

In the model considered in the present paper, the time evolutions of the diagonal and off-diagonal elements of the reduced density operator for the two-level system are decoupled if the initial condition is a product state of the two-level system and the Gibbs state of the reservoir.
Since the work extraction processes in QAPs represented by Eqs.~\eqref{eq:work_1} and \eqref{eq:work_2} depend only on the diagonal elements, we have not included the time evolutions of the off-diagonal elements.
Though the decoupling holds for a wide range of systems with an arbitrary transversal system--environment interaction, i.e., $H_{\rm{SE}}=M_{\rm{S}}\otimes B_{\rm{S}}$ with ${\rm Tr}_{\rm{S}}[\sigma_{z}M_{\rm{S}}]=0$, the quantum coherence represented by the off-diagonal elements may contribute to  further enhancement of the work extraction.
A study of the issue is left for a future investigation.

Finally, we rely on quantum adiabaticity of the work-extracting processes in Eqs.~\eqref{eq:work_1} and \eqref{eq:work_2}, which enables us to expect the maximum work.
However, this assumption also requires an infinitely long time for the processes.
To consider finite-time operations throughout the Otto engine cycle, counter-adiabatic driving \cite{Funo} and shortcut-to-adiabaticity \cite{abah20} have been intensively studied to attempt a concrete realization for experimental studies.
The extension of this work to include non-adiabaticity and the evaluation of the efficiency remains as future work.

\begin{acknowledgments}
This work was supported by JSPS KAKENHI Grant Nos. 15K05200, 15K05207, and 26400409, the Suzuki Foundation, and  partially supported by Grant-in-Aid for Scientific Research on Innovative Areas, Science of Hybrid Quantum Systems, Grant No. 18H04290 and Japan Society for the Promotion of Science KAKENHI Grant No. 19K14611.
\end{acknowledgments}

\appendix
\section{Solution of the time-convolutionless master equation}
\label{appA}
In this appendix, we summarize the solution of the time-convolutionless (TCL) master equation for the spin--boson model~\cite{Guarnieri,Uchiyama14}, which describes the non-Markovian dynamics of the two-level system contacting with the $\mu$th reservoir.

In order to achieve the limit cycle Eq.~(3), we consider the time evolution of the reduced density matrix $\rho^{\mu}_{m}(t)$, which is initially prepared in the state $\ket{m}\!\bra{m}\otimes\rho^{\mu}_{\mathrm{B}}$, where $m=0$ or $1$, and evolves in contact with the hotter (h) or colder (c) reservoir.
The time evolution of $\rho^{\mu}_{m}(t)$ is described by the TCL master equation
	\begin{equation}\label{eq:TCL}
	\frac{\partial}{\partial t}\rho^{\mu}_{m}(t)=\xi^{\mu}(t)\rho^{\mu}_{m}(t),
	\end{equation}
where $\{\mu=\mathrm{h,c}\}$ and $\xi^{\mu}(t)$ is a super-operator called TCL generator.
Up to the second-order cumulant of the system--reservoir coupling, it is given by
	\begin{align}\label{eq:tclgenerator}
	\xi^{\mu}(t)\rho^{\mu}_{m}(t)
&=-i[\mathcal{H}_{\mathrm{S}}^{\mu},\rho^{\mu}_{m}(t)]
\nonumber\\
	&-\int^{t}_{0}d\tau\mathrm{Tr}_{\mu}[\mathcal{H}_{\mathrm{I}}^{\mu},[\breve{\mathcal{H}}_{\mathrm{I}}^{\mu}(-\tau),\rho^{\mu}_{m}(t)\otimes\rho^{\mathrm{eq}}_{\mu}]],
	\end{align}
where $\mathcal{H}_{\mathrm{S}}^{\mu}\equiv\omega_{\mu}\sigma_{z}/2$ is the system Hamiltonian during the contact with hotter/colder reservoir, $\mathrm{Tr}_{\mu}$ stands for a partial trace taken over the $\mu$th reservoir, $\breve{\mathcal{H}}_{\mathrm{I}}^{\mu}(t)\equiv\exp\left[+i(\mathcal{H}_{\mathrm{S}}^{\mu}+\mathcal{H}_{\mathrm{B}}^{\mu})t\right]\mathcal{H}_{\mathrm{I}}^{\mu}\exp\left[-i(\mathcal{H}_{\mathrm{S}}^{\mu}+\mathcal{H}_{\mathrm{B}}^{\mu})t\right]$ is the interaction representation of the coupling Hamiltonian, and $\rho^{\mathrm{eq}}_{\mu}\equiv\exp(-\beta_{\mu}\mathcal{H}_{\mathrm{B}})/\mathrm{Tr}_{\mu}[\exp(-\beta_{\mu}\mathcal{H}_{\mathrm{B}})]$ is the Gibbs state of the $\mu$th reservoir with inverse temperature $\beta_{\mu}\equiv1/k_{B}T_{\mu}$.

Assuming that the system--reservoir coupling is described by the Ohmic spectral density with an exponential cutoff $J(\omega)\equiv\sum_k|g_k|^2\delta(\omega-\epsilon_k) = \lambda\omega\exp(-\omega/\Omega)$, the solution of the TCL master equation Eq.~\eqref{eq:TCL} is given by \cite{Guarnieri},
\begin{align}
	\rho^{\mu}_{m}(t) &= e^{\int_{0}^{t} a^{\mu}(\tau)d\tau}\left(\rho^{\mu}_{m}(0) - \int_{0}^{t}d\tau b^{\mu}(\tau)e^{-\int_{0}^{\tau}a^{\mu}(s)ds}\right),
	\label{eq:solution}
	\end{align}
where
$a^{\mu}(t)\equiv -2\int_{0}^{t} D_{1}^{\mu}(\tau)\mathrm{cos}(\omega_{\mu}\tau)d\tau$,
$b^{\mu}(t)\equiv a^{\mu}(t)/2 -\int_{0}^{t} D_{2}(\tau)\sin(\omega_{\mu}\tau)d\tau$.
The noise and dissipation kernels are given by
\begin{eqnarray}\label{eq:dkernel}
D_{1}^{\mu}(\tau)&&\equiv 2\int_{0}^{\infty}d\omega J(\omega)\coth\left(\frac{\omega}{2T_{\mu}}\right)\cos(\omega\tau)\nonumber\\
&&=2\lambda\biggr(\Omega^{2}\frac{(\Omega\tau)^2-1}{[1+(\Omega\tau)^2]^{2}}
\nonumber\\
&&\phantom{=2\lambda}
+2{T_{\mu}}^2{\rm Re}\biggr\{\psi'\biggr[\frac{T_{\mu}(1+i\Omega\tau)}{\Omega}\biggr]\biggr\}\biggr),
\end{eqnarray}
where $\psi'(z)$ is the derivative of the Euler digamma function $\psi(z)\equiv\Gamma'(z)/\Gamma(z)$, and
\begin{equation}
D_{2}(\tau)\equiv 2\int_{0}^{\infty}d\omega J(\omega)\sin(\omega\tau)=\frac{4\lambda\Omega^3\tau}{[1+(\Omega\tau)^2]^2}.
\end{equation}
Because the second term in the last expression in Eq.~\eqref{eq:dkernel} is proportional to ${T_{\mu}}^2$, the dissipation kernel $D_{1}^{\mu}(\tau)$ may take a large value when $T_{\mu}$ is large, which eventually violates the positivity of the dynamical map for the open-system dynamics even in the weak-coupling regime.
In the present paper, we have carefully chosen parameter values to guarantee the positivity in performing numerical calculations.

\section{Work extraction in the Born--Markov approximation}
\label{appB}
In this appendix, we consider the work extraction in the Born--Markov approximation and show that the quantum Otto engine (QOE) under the Markovian dynamics cannot exceed the Carnot efficiency~\cite{Zhang08,Wang09}.

The Born--Markov approximation is accomplished by taking the long-time (Markovian) limit $t\to\infty$ on the TCL generator Eq.~\eqref{eq:tclgenerator}, whose solution is given by Ref.~\cite{Guarnieri}:
	\begin{widetext}
	\begin{align}
	&\rho^{\mu}_{m,00}(t)= \frac{1+n(\omega_{\mu})}{1+2n(\omega_{\mu})} + \left[\rho^{\mu}_{m,00}(0) - \frac{1+n(\omega_{\mu})}{1+2n(\omega_{\mu})}\right]\exp\left[-2\pi J(\omega_{\mu})\left(1+2n(\omega_{\mu})\right)t\right],\label{eq:Msol1}\\
	&\rho^{\mr{\mu}}_{m,11}(t)=1-\rho^{\mr{\mu}}_{m,00}(t),
	\label{eq:Msol2}
	\end{align}
\end{widetext}
with $n(\omega_{\mu}) = \left(\exp(\omega_{\mu}/T_{\mr{\mu}} )-1\right)^{-1}$.
By definition, the $(0,0)$ components of the initial states are given by $\rho^{\mu}_{1,00}(0)=0$ and $\rho^{\mu}_{0,00}(0)=1$.
We thereby analyze a condition of positive work extraction in the QOE under the Markovian dynamics.
By using Eqs.~(5)--(6), the condition
	\begin{align}
	\ep{W}=\ep{W_1}-\ep{W_2}>0,
	\label{eq:positivework}
	\end{align}
is followed by
\begin{widetext}
	\begin{align}
	(\omega_{\mr{h}}-\omega_{\mr{c}})\left\{\left[P^\mr{h}\rho^\mr{h}_{0,11}(t_{1})+(1-P^\mr{h})\rho^\mr{h}_{0,11}(t_{1})\right]-\left[P^\mr{c}\rho^\mr{c}_{0,11}(t_{2})+(1-P\mr{c})\rho^\mr{c}_{0,11}(t_{2})\right]\right\}>0.
	\end{align}
Since $\omega_{\mr{h}}>\omega_{\mr{c}}$ by definition, we have
	\begin{align}
	\left[P^\mr{h}\rho^\mr{h}_{0,11}(t_{1})+(1-P^\mr{h})\rho^\mr{h}_{0,11}(t_{1})\right]-\left[P^\mr{c}\rho^\mr{c}_{0,11}(t_{2})+(1-P\mr{c})\rho^\mr{c}_{0,11}(t_{2})\right]>0.
	\end{align}
Inserting the solutions Eqs.~\eqref{eq:Msol1}--\eqref{eq:Msol2} to its left-hand side (l.h.s.), we have
	\begin{align}
	\left(\frac{1+n(\omega_{\mr{h}})}{1+2n(\omega_{\mr{h}})}-\frac{1+n(\omega_{\mr{c}})}{1+2n(\omega_{\mr{c}})}\right)\left(1-e^{-2\pi J(\omega_{\mr{h}})(1+2n(\omega_{\mr{h}}))t_{1}}\right)\left(1-e^{-2\pi J(\omega_{\mr{c}})(1+2n(\omega_{\mr{c}}))t_{2}}\right)>0.
	\end{align}
	\end{widetext}
Because the second and third factors in l.h.s. are always positive, we obtain
	\begin{align}
	\frac{1+n(\omega_{\mr{h}})}{1+2n(\omega_{\mr{h}})}>\frac{1+n(\omega_{\mr{c}})}{1+2n(\omega_{\mr{c}})},
	\end{align}
which is transformed to
	\begin{align}
	n(\omega_{\mr{h}})>n(\omega_{\mr{c}}).
	\end{align}
Since $n(\omega_{\mu}) = \left(\exp(\omega_{\mu}/T_{\mu} )-1\right)^{-1}$, we thus show the equivalence of the condition of positive work extraction Eq.~\eqref{eq:positivework} with
	\begin{align}
	\frac{\omega_{\mr{c}}}{\omega_{\mr{h}}}>\frac{T_\mr{c}}{T_\mr{h}},
	\end{align}
and hence
	\begin{align}
	\eta_\mr{O}<\eta_\mr{C}.
	\label{eqB:PWC}
	\end{align}
In short, the Markovian QOE cannot exceed the Carnot efficiency while maintaining positive work extraction.

\begin{figure}[h]
\centering
\includegraphics[clip,width=1 \columnwidth]{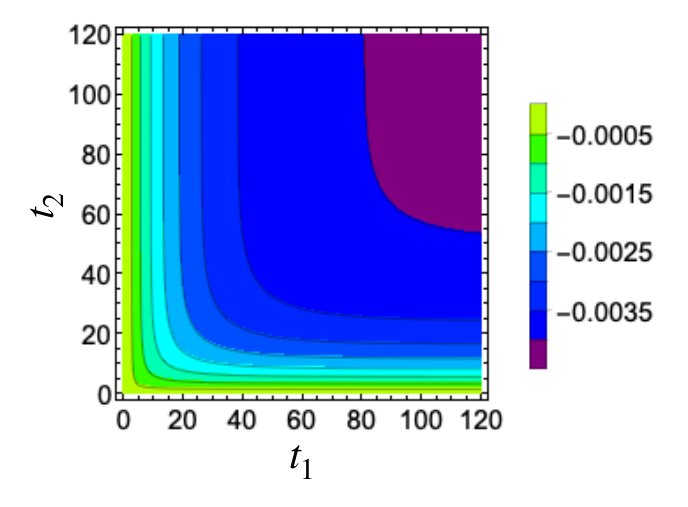}
\caption{\label{fig:Markov} The work extraction from QOE with isochoric processes evaluated under the Born-Markov approximation for the contact durations $t_{1}$ and $t_{2}$ with hotter and colder reservoirs, respectively. We set the parameters to $T_{\mr{h}} = 5.0$, $\lambda = 0.01$, $\Omega = 0.4$, and $\omega_{\mr{c}}/\omega_{\mr{h}} = 0.18$ under the unit of $\omega_{\mr{h}} = 1$ with $k_\mathrm{B}=\hbar=1$, for which $\eta_\mr{O}=0.82>\eta_\mr{C}=0.8$.}
\label{fig2}
\end{figure}
We can understand the relation~\eqref{eqB:PWC} as follows.
It might appear that we could make $\eta_\mr{O}=1-\omega_{\mr{c}}/\omega_{\mr{h}}$ arbitrarily high by adjusting the frequencies $\omega_{\mr{h}}$ and $\omega_{\mr{c}}$ accordingly.
For a very large value of $\omega_{\mr{h}}$, however, the effective temperature~(4) would become higher than $T_\mr{h}$ during the isentropic compression, and hence the system would not be able to receive heat from the hotter reservoir.
Similarly, for a very small value of $\omega_{\mr{c}}$, the system would not be able to dispose heat to the colder reservoir.
In either case, the engine would not function properly, and we would not harvest a positive work, which is exemplified in Fig.~\ref{fig2}.

\section{Work extraction with the definition Eq.~\eqref{eq:work1}}
\label{app:W1}
\begin{figure}
\centering
\includegraphics[clip,width=1\columnwidth]{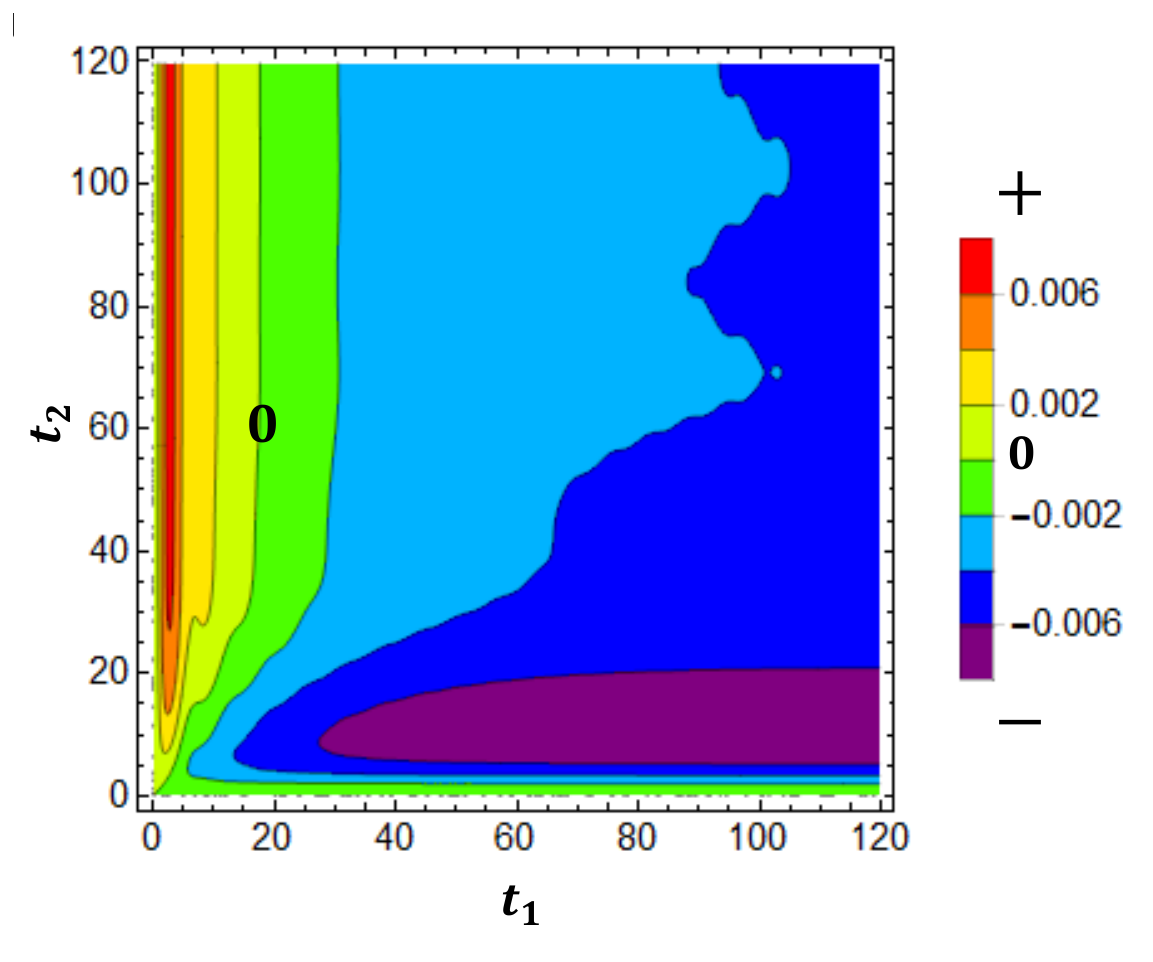}
\vspace{-10pt}
\caption{\label{fig:W1}
Work extracted $W_{\rm{I}}$ of the QOE under the non-Markovian dynamics for contact durations $t_{1}$ and $t_{2}$ evaluated by using the first definition of work Eq.~\eqref{eq:work1}.
For settings corresponding to $\eta_\mr{O}>\eta_\mr{C}$, we find a regime $t_1\lesssim 5$ where $\braket{W}$ is positive.
The parameter settings are: $T_{\mr{h}} = 5.0$, $T_{\mr{c}} = 1.0$, $\lambda = 0.01$, $\Omega = 0.4$, and $\omega_{\mathrm{c}}/\omega_{\mathrm{h}} = 0.18$ for unit values $\omega_{\mathrm{h}} = 1$ with $k_\mathrm{B}=\hbar=1$ (same as in Fig.~\ref{fig:energy_changes}).
} 
\end{figure}
In this appendix, we examine work extraction evaluated with the definition Eq.~\eqref{eq:work1}, excluding the system--reservoir interaction.
In Fig.~\ref{fig:W1}, we present numerical estimates of the extracted work with respect to contact durations $t_{1}$ and $t_{2}$.
We find that the work $W_{\rm{I}}$ becomes positive in the region $t_{1}\lesssim5$, under conditions $\eta_{\rm{O}}>\eta_{\rm{C}}$, which apparently contradicts the Carnot theorem.

The positiveness of the extracted work $W_{\rm{I}}$ for short $t_{1}$ can be understand from temporal changes of the energy flow between the system and the hot reservoir represented by Fig.~\ref{fig:energy_changes}(a) as well as of the effective temperature of the system during the 1st QIP represented by Fig.~\ref{fig:energy_changes}(b).
In the non-Markovian case, by contrast, the energy temporarily flows in reverse, which we call the energy backflow.
Accordingly, the population in the excited state becomes larger as indicated by the increase of the effective temperature of the system shown in Fig.~\ref{fig:energy_changes}(b), which contributes to an increase in the extracted work $W_{\rm{I}}$.
In addition to the backflow, the system also withdraw energy from the system--reservoir interaction in the non-Markovian case, which is indicated by the negative value of the energy change of the interaction presented in Fig.~\ref{fig:energy_changes}(c) (the green solid line).
It may also contributes to a further increase in the work extraction.

\section{Energy change of the hotter reservoir and the two-level system}
\label{appC}
In this appendix we formulate the energy change of the hotter reservoir and the two-level system during the first quantum isochoric process (1st QIP).

Let us first formulate the energy change of the reservoir in terms of the full-counting statistics (FCS) based on the two-point projective measurement.
It is accomplished by successive projective measurements of the reservoir Hamiltonian $\mc{H}^\mr{h}_\mr{B}$.
The measurement scheme is as follows:
first at $t=0$, we perform a measurement of the $\mc{H}^\mr{h}_\mr{B}$ to obtain an outcome $E^\mr{h}_{\mr{B},0}$.
During $0\leq t\leq t_1$, the system undergoes a unitary time evolution brought about by interaction between the system and reservoir.
At $t=t_{1}$, we perform another measurement of $\mc{H}^\mr{h}_\mr{B}$ to obtain another outcome $E^\mr{h}_{\mr{B},t_1}$.
The net energy change of the reservoir during the time interval $t_{1}$ is therefore given by $\Delta E^\mr{h}_{\mr{B}}=E^\mr{h}_{\mr{B},t_1}-E^\mr{h}_{\mr{B},0}$.
The cumulants of $\Delta E^\mr{h}_{\mr{B}}$ are provided by its cumulant generating function 
	\begin{align}\label{eq:cgf}
	S(\chi,t)\equiv\ln\int^{\infty}_{-\infty}P_{t}(\Delta E^\mr{h}_{\mr{B}}) 
	e^{i\chi\Delta E^\mr{h}_{\mr{B}}}d\Delta E^\mr{h}_{\mr{B}},
	\end{align}
where $P_{t}(\Delta E^\mr{h}_{\mr{B}})$ is the probability distribution function of $\Delta E^\mr{h}_{\mr{B}}$ and $\chi$ is the counting field associated with $\mc{H}^\mr{h}_\mr{B}$.
Hence, the expectation value of the energy change during the time interval $t_{1}$ may be expressed by the first derivative of the cumulant generating function,
\begin{equation}\label{eq:mean}
	\langle\Delta\mc{H}^\mr{h}_{\mr{B}}(t_{1})\rangle
	=-\frac{\partial S(\chi,t_{1})}{\partial(i\chi)}\biggr|_{\chi=0}.
	\end{equation}
	
The FCS provides a systematic method of evaluating the cumulant generating function \cite{Esposito}.
Let us formally rewrite it as
	\begin{equation}\label{eq:cgs_densitymatrix}
	S(\chi,t)=\ln{\rm Tr}_{{\rm S}}[\rho^\mr{h}_{\chi}(t)],
	\end{equation}
with
	\begin{equation}
	\rho^\mr{h}_{\chi}(t)\equiv\mr{Tr}_{\mr{h}}[U_{\chi/2}(t,0)W(0)U^{\dagger}_{\chi/2}(t,0)],
	\end{equation}
where $U_{\chi/2}(t,0)\equiv e^{i(\chi/2)\mc{H}^\mr{h}_\mr{B}}U(t,0)e^{-i(\chi/2)\mc{H}^\mr{h}_\mr{B}}$, $U(t,0)$ is the time evolution operator for the total system, and $W(0)$ is the density matrix for the total system at $t=0$.
Assuming a factorized initial condition, the time evolution of the operator $\rho^{(\chi)}(t)$ is described by the equation
\begin{equation}\label{eq:qme}
	\frac{d}{dt}\rho^\mr{h}_{\chi}(t)
	=\xi^{\mr{h}}_{\chi}(t)\rho^\mr{h}_{\chi}(t),
	\end{equation}
which is the TCL-type quantum master equation modified to include the counting field \cite{Uchiyama14}.
Up to the second-order cumulant of the system--reservoir coupling, the TCL generator is given by
	\begin{widetext}
	\begin{equation}\label{secondorder}
	\xi^{\mr{h}}_{\chi}(t)\rho^\mr{h}_{\chi}(t)=-i[\mathcal{H}_{\mathrm{S}}^{\mr{h}},\rho^\mr{h}_{\chi}(t)]-\int^{t}_{0}d\tau\mathrm{Tr}_{\mr{h}}[\mathcal{H}_{\mathrm{I}}^{\mr{h}},[\breve{\mathcal{H}}_{\mathrm{I}}^{\mr{h}}(-\tau),\rho^\mr{h}_{\chi}(t)\otimes\rho^{\mathrm{eq}}_{\mr{h}}]_{\chi}]_{\chi},
	\end{equation}
where $[X,Y]_{\chi}\equiv X^{(\chi)}Y-YX^{(-\chi)}$ with $X^{(\chi)}\equiv e^{i\chi\mc{H}^\mr{h}_\mr{B}/2}Xe^{-i\chi\mc{H}^\mr{h}_\mr{B}/2}$.
We note that the familiar master equation describing the time evolution of the usual density operator is recovered by taking $\chi=0$ on Eq.~(\ref{eq:qme}).
In terms of the TCL master equation formalism, the mean dissipated heat is expressed by \cite{Uchiyama14}
	\begin{equation}\label{eq:heat}
	\langle\Delta\mc{H}^\mr{h}_{\mr{B}}(t_{1})\rangle=-\int^{t_{1}}_{0}dt{\rm Tr}_{{\rm S}}\Biggr[
	\frac{\partial\xi^\mr{h}_{\chi}(t)}{\partial(i\chi)}\biggr|_{\eta=0}
	\rho^\mr{h}(t)\Biggr].
	\end{equation}
By applying the expression Eq.~\eqref{eq:heat} to the spin--boson model, we obtain the expression of the energy change of the hotter reservoir during the 1st QIP,
\begin{align}
	\left<\Delta \mathcal{H}_{\mathrm{B}}^{\mr{h}}(t_1)\right> = &\omega_{h}P^\mathrm{h}(\rho_{0,00}^{\mr{h}}(t_1)-1)+\omega_{\mr{h}}(1-P^\mr{h})\rho_{1,00}^{\mr{h}}(t_{1}) - 1\nonumber\\
	&+ \int_{0}^{t_1}d\tau\left\{
P^\mr{h}\left[\left(2\rho_{0,00}^{\mr{h}}(\tau)-1\right)
D_{1}^{\mr{h}}(\tau)\sin(\omega_{\mr{h}}\tau)+D_{2}(\tau)\cos(\omega_{\mr{h}}\tau)\right]\right.
\nonumber\\
 	&\qquad\qquad+\left. (1-P^\mr{h})\left[\left(2\rho_{1,00}^{\mr{h}}(\tau)-1\right)D_{1}^{\mr{h}}(\tau)\sin(\omega_{\mr{h}}\tau)+D_{2}(\tau)\cos(\omega_{\mr{h}}\tau)\right]\right\}.
	\label{eq:bathenergy}
	\end{align}
We next formulate the energy change of the two-level system.
Because the energy level of the two level system is unchanged during the QIP, the net energy change of the system is simply evaluated as
	\begin{align}
	\left<\Delta \mathcal{H}_{\mr{S}}^{\mr{h}}(t_1)\right>&=\omega_{\mr{h}}(\rho^\mr{h}_{11}(t_1)-\rho^\mr{h}_{11}(0))=\omega_{\mr{h}}P^\mathrm{h}\rho_{0,11}^{\mr{h}}(t_1)+\omega_{\mr{h}}(1-P^\mr{h})\left(\rho_{1,11}^{\mr{h}}(t_{1}) - 1\right).
	\end{align}
Using these expressions with unity of the total population $\mathop{\mathrm{Tr}} \rho_{0(1)}(t)=\rho_{0(1),00}(t)+\rho_{0(1),11}(t)=1$, we find the change in the interaction energy as
	\begin{align}\label{eq:changeint}
	\left<\Delta \mathcal{H}_{\mr{I}}^{\mr{h}}(t_1)\right>=& \int_{0}^{t_1}d\tau\left\{
P^\mr{h}\left[\left(2\rho_{0,00}^{\mr{h}}(\tau)-1\right)
D_{1}^{\mr{h}}(\tau)\sin(\omega_{\mr{h}}\tau)+D_{2}(\tau)\cos(\omega_{\mr{h}}\tau)\right]\right.
\nonumber\\
 	&\qquad\qquad+\left. (1-P^\mr{h})\left[\left(2\rho_{1,00}^{\mr{h}}(\tau)-1\right)D_{1}^{\mr{h}}(\tau)\sin(\omega_{\mr{h}}\tau)+D_{2}(\tau)\cos(\omega_{\mr{h}}\tau)\right]\right\}.
	\end{align}
\end{widetext}

\section{Dependence of work extraction on $\omega_{\mr{h}}$ and $\omega_{\mr{c}}$}
\label{appD}

\begin{figure*}[h]
\centering
\includegraphics[clip,width=0.7 \textwidth]{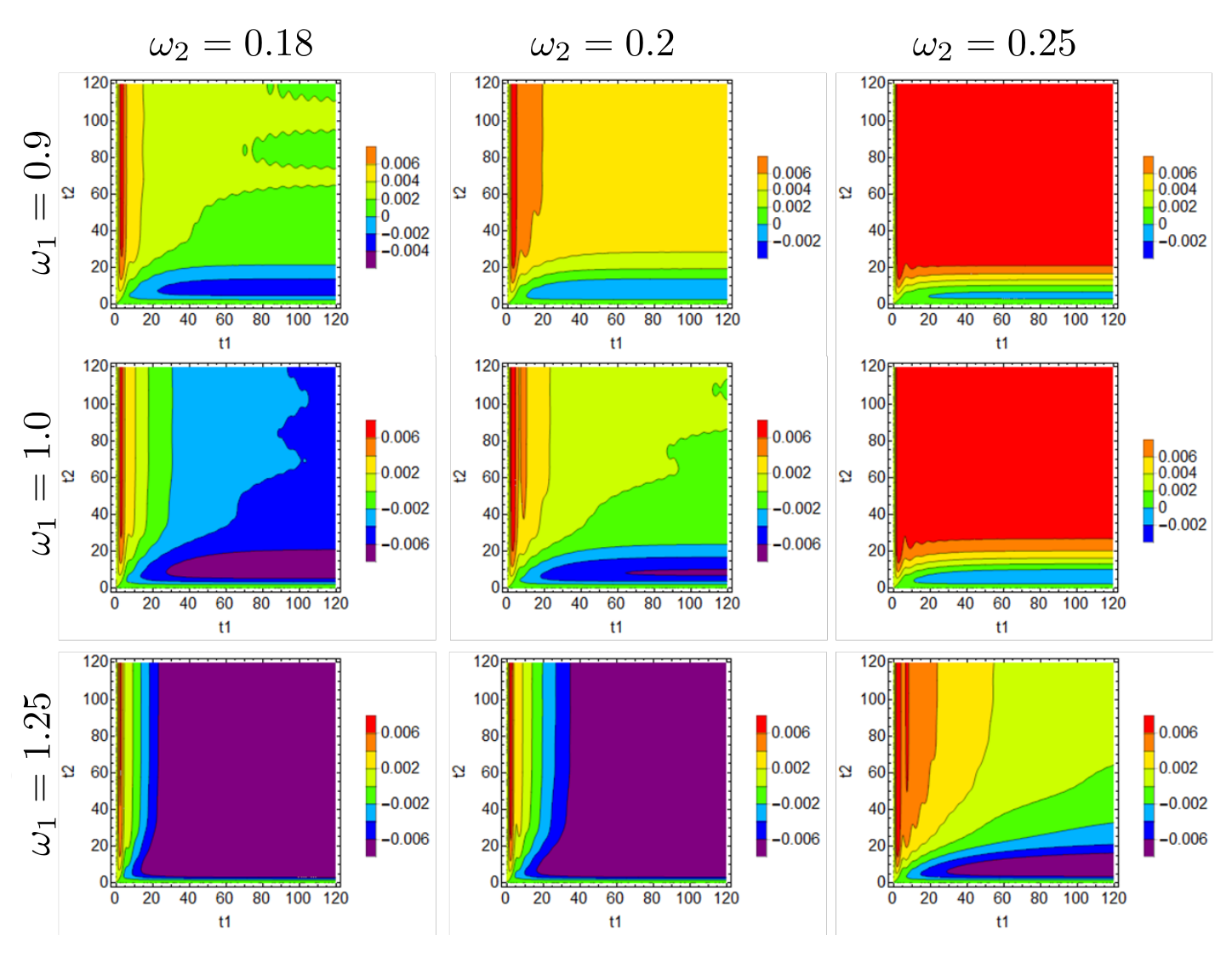}
\caption{Dependence of the work extraction on $\omega_{\mr{h}}$ and $\omega_{\mr{c}}$ evaluated under non-Markovian dynamics for the contact durations $t_{1}$ and $t_{2}$ with hotter and colder reservoirs, respectively. We set the parameters to $T_{\mr{h}} = 5.0$, $\lambda = 0.01$, $\Omega = 0.4$ under the unit of $k_{{\rm B}}=\hbar=1$.}
\label{fig:supplfig2}
\end{figure*}
In this appendix, we summarize the dependence of the work extraction on several $\omega_{\mr{h}}$ and $\omega_{\mr{c}}$.
In Fig.~\ref{fig:supplfig2}, we provide work extractions evaluated by means of Eqs.~(3)--(4) in the main text under the non-Markovian dynamics for several combinations of $\omega_{\mr{h}}$ and $\omega_{\mr{c}}$.
In the figure, the three panels on the diagonal line from upper left to lower left ($(\omega_{\mr{h}},\omega_{\mr{c}})=(0.9,0.18),(1.0,0.2),(1.25,0.25)$) are evaluated under the condition $\eta_{\mathrm{O}}=\eta_{\mathrm{C}}=0.8$.
The three panels above the diagonal line ($(\omega_{\mr{h}},\omega_{\mr{c}})=(0.9,0.2),(0.9,0.25),(1.0,0.25)$) correspond to $\eta_{{\rm O}}<\eta_{{\rm C}}$, and the other three panels ($(\omega_{\mr{h}},\omega_{\mr{c}})=(1.0,0.18),(1.25,0.18),(1.25,0.2)$) correspond to $\eta_{{\rm O}}>\eta_{{\rm C}}$.
The figure shows that, under the condition $\eta_{{\rm O}}\leq\eta_{{\rm C}}$, we can extract positive work form the QOE for a wide range of contact durations $(t_{1},t_{2})$.
For $\eta_{{\rm O}}\leq\eta_{{\rm C}}$, in contrast, the work extraction is negative for the majority of $(t_{1},t_{2})$, but we can still extract positive work if the contact duration wit the hotter bath, $t_{1}$, is sufficiently short.

\begin{figure*}[h]
\centering
\includegraphics[clip,width=0.7 \textwidth]{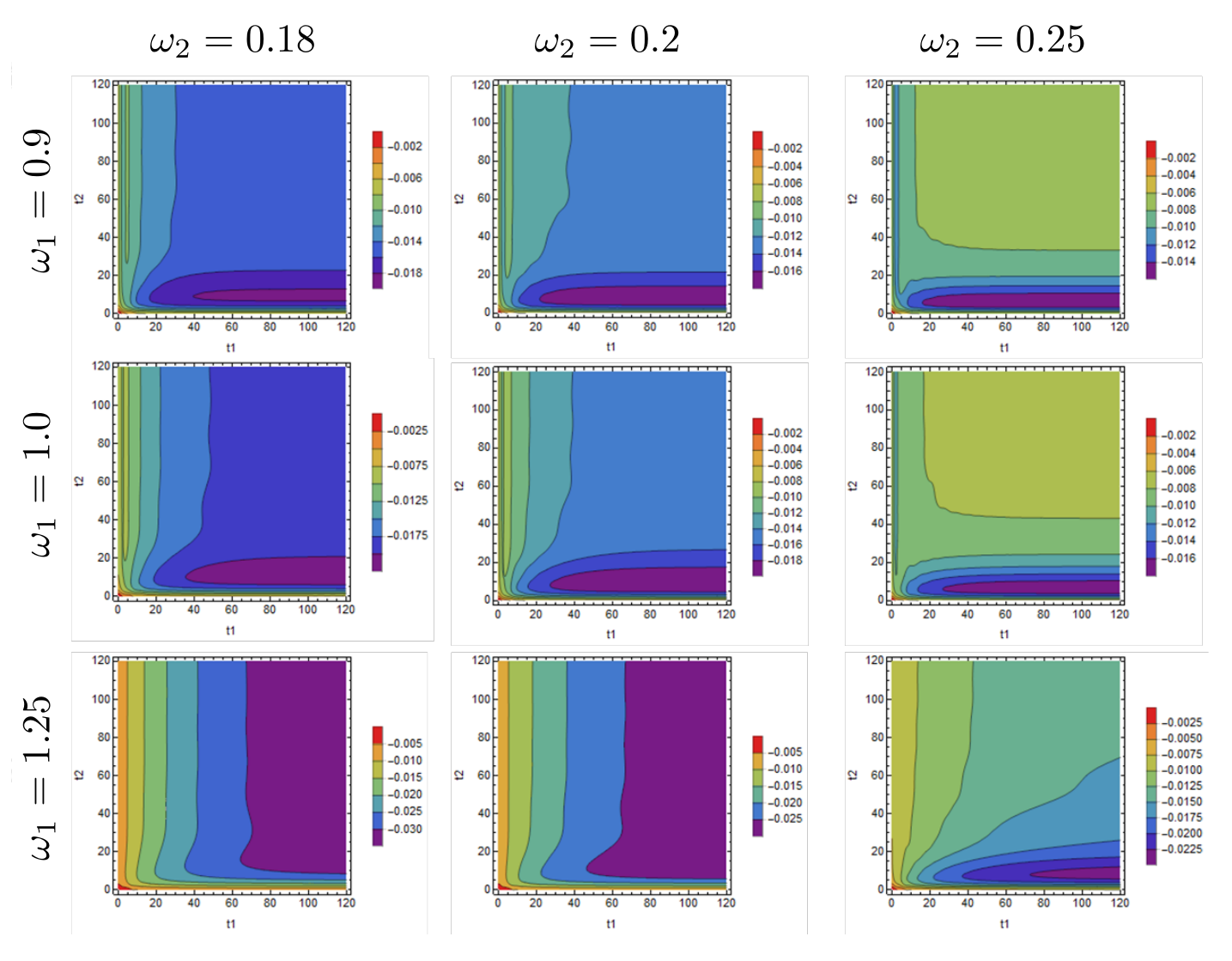}
\caption{Dependence of the work extraction on $\omega_{\mr{h}}$ and $\omega_{\mr{c}}$ evaluated under non-Markovian dynamics for the contact duration $t_{1}$ and $t_{2}$ with the detachment energy counted in. We set the parameters to $T_{\mr{h}} = 5.0$, $\lambda = 0.01$, $\Omega = 0.4$ under the unit of $k_{{\rm B}}=\hbar=1$.}
\label{fig:supplfig3}
\end{figure*}
In Fig.~\ref{fig:supplfig3}, we also provide work extractions for several combinations of $\omega_{\mr{h}}$ and $\omega_{\mr{c}}$ with the detachment energy counted in as a part of work.
In this figure, we find that the work extraction is always negative for any combinations of $(\omega_{\mr{h}},\omega_{\mr{c}})$.
This means that the efficiency of the quantum Otto engine cannot exceed the Carnot efficiency if we include the detachment energy as a part of work.

\end{document}